# Applying Large Language Models to Travel Satisfaction Analysis

Pengfei Xu

*Department of Geography, Hong Kong Baptist University,*

*Kowloon Tong, Kowloon, Hong Kong.*

*Email: xpfovo@life.hkbu.edu.hk*

Donggen Wang (corresponding author)

*Department of Geography, Hong Kong Baptist University,*

*Kowloon Tong, Kowloon, Hong Kong.*

*Email: dgwang@hkbu.edu.hk*

## Abstract

As a specific domain of subjective well-being, travel satisfaction has recently attracted much research attention. Previous studies primarily relied on statistical models and, more recently, machine learning models to explore its determinants. Both approaches, however, depend on sufficiently large sample sizes and appropriate statistical assumptions. The emergence of Large Language Models (LLMs) offers a new modeling approach that can address these limitations. Pre-trained on extensive datasets, LLMs have strong capabilities in contextual understanding and generalization, significantly reducing their dependence on task-specific data and stringent statistical assumptions. The main challenge in applying LLMs lies in the behavioral misalignment between LLMs and humans. Using household survey data collected in Shanghai, this study identifies the existence and source of misalignment, and applies a few-shot learning method to address the misalignment issue. We find that the zero-shot LLM exhibits behavioral misalignment, leading to low prediction accuracy. With just a few samples, few-shot learning can align LLMs and enable them to outperform baseline models. Discrepancies in variable importance among machine learning model, zero-shot LLM, and few-shot LLM reveal that the misalignment arises from the gap between the general knowledge embedded in pre-trained LLMs and the specific, unique characteristics of the dataset. On these bases, we propose an LLM-based modeling approach that can be applied to model travel behavior with small sample sizes. This study highlights the potential of LLMs for modeling not only travel satisfaction but also broader aspects of travel behavior.

## 1. Introduction

Travel satisfaction evaluates individuals' daily travel experience from both affective and cognitive perspectives, serving as a key indicator of the quality of transport services (Ettema et al., 2010). Numerous empirical studies have explored the determinants of travel satisfaction, identifying significant influences from socioeconomic factors (De Vos et al., 2021), the built environment (Chen eta al., 2024; G. Liu et al., 2024), and travel characteristics (De Vos et al., 2019; St-Louis et al., 2014). More recent research has focused on the conceptualization and evaluation mechanism of travel satisfaction, revealing that factors such as peer's travel, travel experience, travel captivity, and travel preferences serve as important reference points in the assessment of travel satisfaction (Guan et al., 2023; Guan and Wang, 2024).

Existing research on travel satisfaction has predominantly relied on statistical models (e.g., De Vos et al., 2019; De Vos and Witlox, 2017; Guan et al., 2023). More recent studies, however, have increasingly leveraged machine learning techniques (e.g., Chen et al., 2024; Dong et al., 2025; G. Liu et al., 2024). Both modeling approaches, especially the latter one, relies on large quantities of data. Statistical models, such as linear regression (LR) and structural equation models (SEM), are based on prior assumptions on the statistical distributions and relationships of variables. For example, both LR and SEM assume that the relationships between variables are linear. The paths or relationships among variables in SEM must be hypothesized based on a thorough review of the existing literatures (Chen et al., 2024; Wang et al., 2025; Zhao et al., 2020). On the other hand, machine learning models, such as random forest (RF) and gradient boosting decision trees (GBDT), are non-parametric and explore data without requiring prior knowledge or assumptions about its structure. However, missing data, which is prevalent in questionnaire surveys, often poses significant challenges for adequately training machine learning models (Dodeen, 2018). Another point of contention is the "black-box" nature of these models. Despite recent advancements in interpretable and explainable machine learning, behaviorally unreasonable effects can still emerge. For instance, Zhao et al. (2020) found that while machine learning models outperform logit models in predicting travel mode choices, the marginal effects of independent variables can be irrational, and such errors are often difficult to identify.

The emergence of large language models (LLMs) has provided new modeling approaches for travel behavior research. Popular pre-trained LLMs, such as DeepSeek-R1 and GPT-4, have demonstrated superior performance in travel mode prediction, mobility generation and prediction (Mo et al., 2023; Wang et al., 2024; Wang et al., 2023). Compared with statistical models and machine learning methods, LLMs have several distinct comparative advantages. Firstly, they are trained on a wide variety of data of both qualitative and quantitative, thus possessing encyclopedic knowledge for any downstream (or domain specific) task even without finetuning (Liang et al., 2024). Secondly, the massive parameters of LLMs enable them to quickly learn novel patterns with minimal example samples (T. Liu et al., 2024). Thirdly, LLMs have powerful semantic understanding capabilities like human beings, enabling them to comprehend complex and lengthy contexts based on appropriate prompts (Mo et al., 2023). Finally, the outputs of LLMs have strong interpretability. This is exemplified by their unique ability of providing the reasoning process in natural language, a capability that machine learning models and earlier deep neural network models are unable to achieve.

Despite the incredible capability of LLMs, a critical challenge to apply LLMs for travel behavior modeling is behavioral misalignment, which refers to the extent to which the output and reasoning process of LLMs are inconsistent with human behavior in values, intentions, preferences, and ethical standards, etc. (Goli and Singh, 2023; Leng, 2024; T. Liu et al., 2024). T. Liu et al. (2024) is among the first, if not the first, to acknowledge and address the misalignment phenomenon of LLMs in travel behavior modeling. They revealed the misalignment in travel mode choice, though they did not examine the causes of misalignment. To contribute to this fast-growing field of LLM-based travel behavior modeling, this study is designed to explore the misalignment issues in applying LLMs for modeling travel satisfaction, identify the source of misalignment, and develop methods to address the issues. We shall answer the following three research questions: firstly, is there also a misalignment between LLMs and human travelers in travel satisfaction? secondly, how to improve their alignment? and lastly, what are the underlying causes of this misalignment?

To this end, we propose a prompting approach to predicting travel satisfaction. Data used for experiments are obtained from a household travel survey conducted in 2018 in Shanghai,

China. We select four mainstream pre-trained LLMs as our experimental models, while LR and GBDT—commonly used methods in travel satisfaction modeling—are employed as baseline models. First, we construct a well-designed prompt to predict individuals' travel satisfaction without any additional guidance. Second, we use few-shot learning to align LLM. The performance of LLM and two baseline models are compared to assess the effectiveness of aligning process. Third, the variable importance in GBDT and LLM are compared, and reasoning contents of LLM are analyzed to investigate the misalignment in depth. Finally, we shall discuss the generalizability of our modeling approaches and the potentials of LLMs for travel behavior modeling in general.

The remainder of this paper is organized as follows. We shall first critically review the related literature on the determinants of travel satisfaction and LLMs' application in travel behavior modeling in Section 2. This is followed by research design and the introduction of our modeling approaches in Section 3. Section 4 presents the experiment results. The last section summarizes the research findings, discusses their implications, and makes recommendations for future research.

## 2. Literature review

### *2.1 Influential factors of travel satisfaction*

Travel satisfaction measures how people are satisfied with their daily travel in general or a specific trip (De Vos and Witlox, 2017). The role of socioeconomics, travel characteristics, and the built environment in travel satisfaction has been examined in previous studies. Many of them have confirmed that gender (De Vos et al., 2016), age, income (Ye and Titheridge, 2017), and education level (De Vos et al., 2021) are significant determinants of travel satisfaction. Regarding household car ownership, De Vos et al. (2021) found that it is negatively associated with commuting satisfaction in Ghent, Belgium, while Li et al. (2022) revealed that its influence on overall travel satisfaction is not significant in Beijing, China. They further found that owning more than one car can improve life satisfaction, whereas no impact on travel satisfaction.

The important role of travel characteristics for travel satisfaction has been frequently acknowledged in the literature. Based on data from a questionnaire survey in Shanghai, Chen et al. (2024) found that commuting characteristics, including duration, transport mode and distance, are the most important factors explaining travel satisfaction. In terms of transport mode, car users are typically more satisfied than public transit users (Mao et al., 2016; Ye and Titheridge, 2017). Guan et al. (2023) argued that the generally low satisfaction with public transit may be related to the service quality itself, rather than merely preference. Active and non-motorized travels are generally considered to have higher travel satisfaction than motorized travels do (De Vos et al., 2016, 2013). It is commonly found that longer commuting durations usually lead to lower travel satisfaction (De Vos et al., 2019; F. Wang et al., 2020). For travel frequency, an empirical study by Waygood et al. (2019) showed that frequent travelers tend to be more unsatisfied than non-frequent travelers for all travel modes except cycling. Some non-instrumental factors, such as in-vehicle activities, cleanness, and noise, are also found to influence satisfaction with bus travel (Carreira et al., 2014; Eriksson et al., 2013; Ettema et al., 2012).

While research on the influence of the built environment on travel behavior has a long-standing history, studies focusing on its impacts on travel satisfaction is relatively scarce. Some studies explored the direct and indirect impacts of the built environment on travel satisfaction: direct impacts enhance travel satisfaction by improving accessibility to facilities or scenic landscape (De Vos and Witlox, 2017), while indirect effects influence travel satisfaction by shaping transport mode choices (Ye and Titheridge, 2017). The accessibility to facilities and the safety of walking are very important factors affecting travel satisfaction. Ettema et al. (2011) found that travel satisfaction is positively associated with the accessibility to public transit. Kim et al. (2014) concluded that the pedestrian crossings and the width of sidewalks have notable impacts on walking satisfaction. Besides, it has been reported that residential relocation-induced changes in the built environment have both direct and indirect influence in travel satisfaction (F. Wang et al., 2020). The linear relationship between variables have been challenged in recent years. Consequently, studies about non-linear relationships between the built environment variables and travel satisfaction have proliferated with the development of interpretable/explainable machine learning methods (G. Liu et al., 2024; Wang et al., 2025). Random Forest and GBDT are representative models in this developing field. Dong et al. (2025)

found that both the number of parking lots and the proximity to subway lines influence the travel satisfaction with non-linear effects. Chen et al. (2024) revealed the non-linearity of the impacts of the built environment on commuting satisfaction, underlining the deficiency of the modeling approaches based on the linear assumption.

Due to the strong subjectivity of well-being, many studies have made efforts to examine the evaluation mechanisms of travel satisfaction from a psychological perspective. Some studies have pointed out that travel satisfaction may be assessed with reference points, the most used of which may be the counterfactually ideal travel. For example, people tend to be more satisfied when travel is consonant (Guan and Wang, 2024) or the ideal travel duration can be guaranteed (Humagain and Singleton, 2020). Inspired by well-known prospect theory (Kahneman and Tversky, 2013), Guan et al. (2023) revealed that travel experience, peers' travel, and travel preference serve as reference points in travel satisfaction evaluation and the influence of these reference points vary across different travel modes. From the perspective of utility theory, De Vos et al. (2016) defined travel satisfaction as an experienced utility, referring to the experience of feelings and emotions. Regarding the pivotal role of travel satisfaction in conceptualizing a travel behavior process, daily travel satisfaction is argued to serve as a middle-term satisfaction influenced by short-term trip satisfaction and long-term life satisfaction (De Vos and Witlox, 2017; Rao et al., 2025).

Although these studies have provided valuable insights into the determinants of travel satisfaction, they predominantly rely on statistical models (e. g., linear regression, structural equation modeling). When incorporating built environment variables, machine learning methods are often employed to capture the non-linear relationship and threshold effect. Both statistical models and machine learning methods are data-driven, while LLMs, trained on extensive qualitative and quantitative data, are equipped with general knowledge from diverse scenarios and datasets. This characteristic endows LLMs the potential to advance our understanding of travel satisfaction.

### *2.2 Large language models in travel behavior modeling*

The rapid advancement of LLMs in recent years has sparked growing interest in their application to travel behavior research. Studies have increasingly utilized LLMs for tasks such as predicting individuals' daily activities and mobility patterns (Wang et al., 2024, 2023), as well as forecasting origin-destination (OD) flows and regional-level travel demand (Liang et al., 2024; Yu et al., 2024; Yuan et al., 2024).

LLMs have been increasingly used in spatial-temporal prediction at a region or city level, aiming to predict the dynamic patterns of urban activities across both space and time. Yuan et al. (2024) designed a novel LLM-based prediction framework named UniST, and employed both pre-training and prompt strategies to enhance its accuracy. Due to the high transferability of pre-trained LLMs, the UniST was proved to perform well across various scenarios such as taxi OD prediction, regional bike usage, etc. Liang et al. (2024) made use of textual information on public events as prompts for LLMs to predict taxi travel demand around an indoor arena. This recent study was the first attempt to use LLMs to predict traffic under non-recurrent events. Yu et al. (2024) transferred origin-destination (OD) forecasts from one city to another and discovered the high transferability of LLMs. They credit this transferability to the LLMs' robust semantic understanding capabilities. It is important to note that earlier deep learning models generally incorporated event information in a structured and numeric form (such as an event matrix), which had limited capability for feature expression (e. g., Fang et al., 2020; Liu et al., 2019; Zhou et al., 2019). The powerful semantic understanding abilities of LLMs expand the possibilities for integrating multi-source data into the prediction process, offering new ways for aggregate-level traffic flow prediction.

LLMs have also been employed to predict individual daily activity and mobility patterns. Wang et al. (2023) integrated individuals' profiles along with the times of historical stays, context stays, and target stays as prompts for LLMs to predict their next activity location. Li et al. (2024) designed a two-stage prompting framework to generate daily activities based on the strong pattern extraction ability of LLMs. Shao et al. (2024) made the first attempt to incorporate the Theory of Planned Behavior as prompts for LLMs to predict an individual's next activity. Given the challenges of obtaining mobility data due to privacy concerns, Y. Liu et al. (2024) demonstrated that it is possible to accurately predict activity chains using only basic socio-demographic information, without relying on historical activity and travel data. Mo et al. (2023) explored the application of LLMs for transport mode choice prediction, using travel characteristics and personal socioeconomics as prompts. Existing studies attributed the remarkable capabilities of LLMs in mobility prediction and generation to several key advantages: (1) their efficiency in understanding both unstructured natural language and structured numerical data (Wang et al., 2024); (2) their ability to summarize historical mobility or travel patterns (Li et al., 2024); and (3) their emergent capabilities—advanced functionalities that are absent in smaller models but present in larger ones (Wang et al., 2023; Wei et al., 2022). Besides these inherent abilities of LLMs, the effective formatting of individual profiles and the carefully designed prompts are also important.

Although the studies mentioned above have demonstrated the superior performance of LLMs at both aggregate and individual levels, most have overlooked two important issues: the misalignment between LLMs and humans in decision making and how to address it. This misalignment may lead to misleading results, as significant differences exist between the choices made by pre-trained LLMs and those made by human decision-makers (Goli and Singh, 2023; Hagendorff et al., 2023; Leng, 2024). Wang et al. (2024) aligned LLMs with real-world urban mobility data through a self-consistency approach and a retrieval-augmented strategy, but they fell short of in-depth investigation on the causes of misalignment and the behavioral changes of LLMs during the aligning process. T. Liu et al. (2024) is one of the few studies that explicitly acknowledges the misalignment between LLMs and human trip-makers by comparing the accuracy of transport mode choice predictions between LLMs and the logit model. While they did not identify the specific sources of this misalignment, they proposed two methods—persona loading and few-shot learning—to mitigate its effects.

Overall, although LLMs has drawn considerable research attention and demonstrated great ability in travel behavior modeling, there is still limited understanding of the misalignment between LLMs and human travelers. For the future application of LLMs in travel behavior modeling, it is important to investigate the source of misalignment and develop strategies to address the issue.

## 3. Methodology

### *3.1 Data*

The data were derived from a household travel survey conducted in Shanghai, China, between August and October 2018, covering all sixteen districts of Shanghai (excluding Chongming Island). A stratified sampling method with probability-proportional-to-size (PPS) method was employed to ensure the comprehensiveness of samples. For the sampled households, face-to-face interviews were conducted, and all family members above 12 years old were required to complete questionnaires, including information on household and personal socio-economics, residential environments, commuting characteristics, and travel satisfaction, etc. In total, 2144 completes from 1064 households completed the questionnaire. Overall, the sample is somewhat biased towards younger, highly educated, and higher-income individuals, and thus may not be fully representative of the broader Shanghai population in terms of socio-demographic characteristics (please refer to Mao et al. (2022) for details). Since not all respondents provided information on the variables that are needed for this study, only 874 respondents are included in our sample.

Travel satisfaction is assessed by the 9-item scale developed by Ettema et al. (2011). This scale contains statements about their perception of daily travel, as listed in Table 1. Respondents rated each item on a scale of 1 to 7, and the average of all items is taken as the travel satisfaction in this study. The constructed prompt includes four dimensions: socio-economics, built environment, travel characteristics, and reference points to comprehensively reflect the objective and perceived determinants of travel satisfaction. Socio-economic variables include gender, age, income, education level, and private car access. The built environment variables include the self-reported walking times from residence to the nearest public transit station, parking lot, shopping mall, hospital, and restaurant. Travel characteristics are indicated by the number of trips on weekday, the daily commuting time, and the commuting mode to capture the effects of travel frequency, travel duration, and travel mode on travel satisfaction. Following Guan et al. (2023), respondent's family members are treated as their peers. Given that residential relocation is often regarded as a milepost event affecting the evaluation of daily travel (Gerber et al., 2017; F. Wang et al., 2020), commuting time and mode before the most recent relocation are treated as past travel experience. A detailed description of the variables that are used for prompt is given in Table 2.

**Table 1** The satisfaction with travel scale

| | |
|---|---|
| *Positive deactivation-negative activation* | |
| I felt time was pressed – relaxed in daily travel | Very hurried (1) – Very relaxed (7) |
| I was worried – confident I would be in time in daily travel | Very worried (1) – Very confident (7) |
| I was stressed – calm in daily travel | Very stressed (1) – Very calm (7) |
| *Positive activation-negative deactivation* | |
| I was tired – alert in daily travel | Very tired (1) – Very alert (7) |
| I was bored – enthusiastic in daily travel | Very bored (1) – Very enthusiastic (7) |
| I was fed up – engaged with daily travel | Very fed up (1) – Very engaged (7) |
| *Cognitive evaluation* | |
| My daily travel was low – high standard | Very low standard (1) – Very high standard (7) |
| My daily travel worked poorly – well | Very poorly (1) – Very well (7) |
| My daily travel is the worst – best I can think of | Worst I can think of (1) –Best I can think of (7) |

**Table 2** Full list of variables used in the prompt and baseline models

| Dimension | Variable | Description | Mean/% |
|---|---|---|---|
| Socioeconomics | Gender | 0: male | 54.71% |
| | | 1: female | 45.29% |
| | Age | Numeric | 34.71 |
| | Income | Numeric (yuan per month) | 21650 |
| | Education level | 1: Primary school | 2.23% |
| | | 2: Middle school | 5.03% |
| | | 3: High school | 7.56% |
| | | 4: Junior college | 7.42% |
| | | 5: Bachelor | 63.71% |
| | | 6: Master or above | 14.04% |
| | Car access | 0: no car | 80.89% |
| | | 1: one car | 1.33% |
| | | 2: two cars | 14.20% |
| | | 3: three cars | 2.47% |
| | | 4: four cars | 0.31% |
| Built environment (Walking time from home to nearest …) | Public transit station | Numeric in minutes | 9.23 |
| | Parking lot | Numeric in minutes | 10.69 |
| | Hospital | Numeric in minutes | 20.50 |
| | Shopping mall | Numeric in minutes | 15.20 |
| | Restaurant | Numeric in minutes | 12.26 |
| Travel characteristics | Commuting time | Numeric in minutes | 26.97 |
| | Commuting mode | 1: walk | 3.18% |
| | | 2: bike | 2.00% |
| | | 3: subway | 24.00% |
| | | 4: taxi | 0.94% |
| | | 5: bus | 15.80% |
| | | 6: private car | 48.75% |
| | | 7: shuttle bus | 1.53% |
| | | 8: car-sharing | 0.41% |
| | | 9: others | 3.39% |
| | Number of trips on weekday | Numeric | 5.27 |
| Reference points | Past commuting time | Numeric in minutes | 27.19 |
| | Past commuting mode | 1: walk | 5.54% |
| | | 2: bike | 4.98% |
| | | 3: subway | 19.77% |
| | | 4: taxi | 2.72% |
| | | 5: bus | 27.01% |
| | | 6: private car | 23.02% |
| | | 7: shuttle bus | 7.98% |
| | | 8: car-sharing | 0.23% |
| | | 9: others | 8.75% |
| | Peer's commuting time | Numeric in minutes | 27.30 |
| | Peer's commuting mode | 1: walk | 1.45% |
| | | 2: bike | 0.83% |
| | | 3: subway | 26.92% |
| | | 4: taxi | 1.24% |
| | | 5: bus | 12.01% |
| | | 6: private car | 52.58% |
| | | 7: shuttle bus | 2.28% |
| | | 8: car-sharing | 0.62% |
| | | 9: others | 0.60% |
| \ | Travel satisfaction | Numeric | 4.38 |

### *3.2 Method*

The primary objective of this study is to address the three research questions listed earlier. We first need to identify the LLMs for our study. To enhance the generalizability of this study, we choose four top-notch pre-trained LLMs: DeepSeek-R1-671b, Qwen-3-235b, GPT-4o-200b, and Llama-3.3-70b, where 'b' denotes the number of parameters in billions. Among them, DeepSeek-R1 is renowned for its unique capability for deep thinking, allowing us to gain valuable insights into its reasoning process and output results. To answer the first research question, we construct a prompt to enable the LLM to predict individuals' travel satisfaction without any guidance, which we refer to as the zero-shot approach, as illustrated in Figure 1. We develop a few-shot prediction framework as illustrated in Figure 2 to address the second research question. The LLM first learns from extremely limited samples and subsequently predicts for the remaining samples. To address the last research question, we identify and examine the differences in variable importance between GBDT and LLMs and the divergences in the reasoning contents between the zero-shot LLM and the few-shot LLM. The outputs of LLMs include travel satisfaction scores and the normalized variable importance. Though LLMs is pre-trained on multi-source data, only a small portion of the data is related to travel behavior modeling, which may lead to a severe misalignment. We can quantify the degree of misalignment by examining the prediction accuracy metrics.

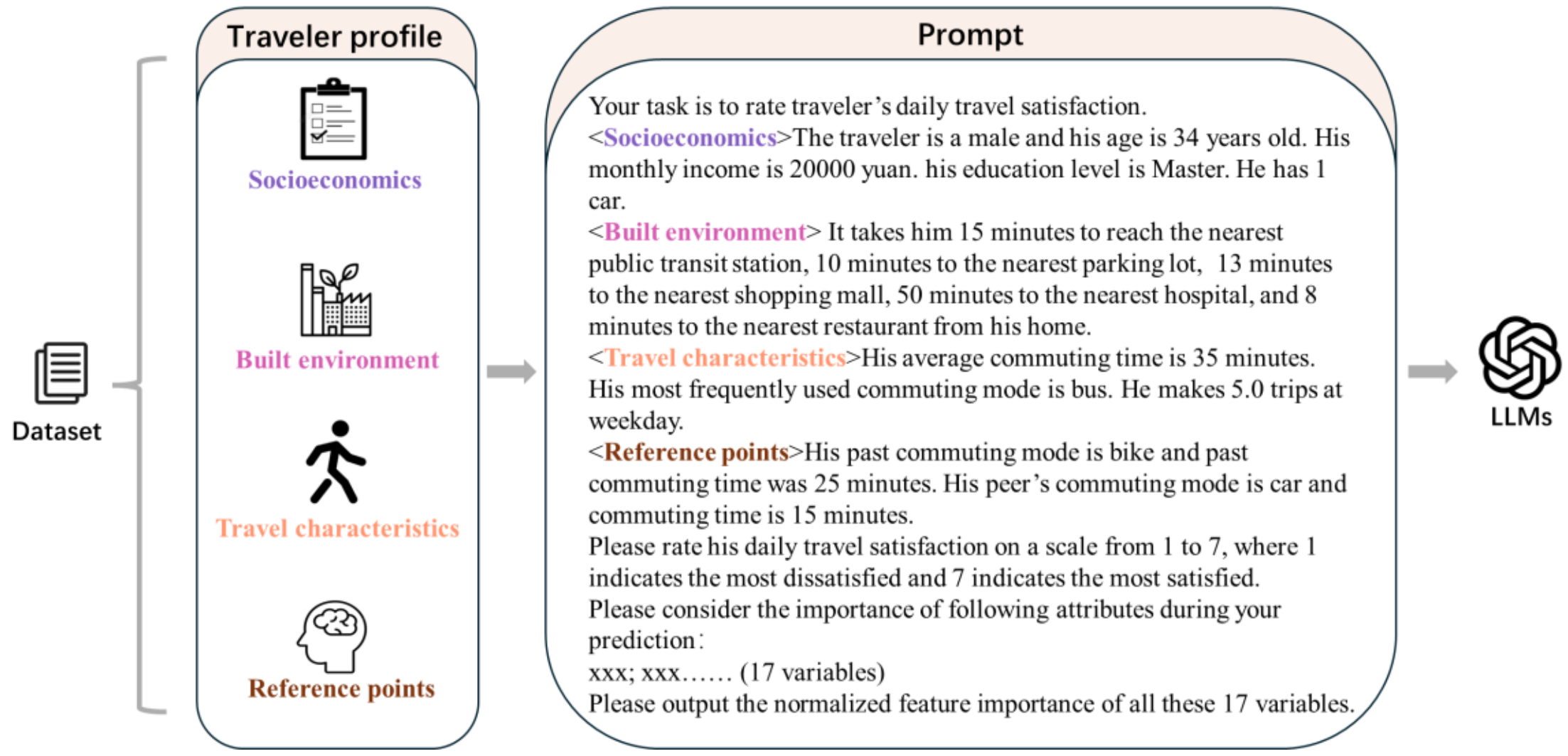

**Figure 1** Zero-shot prediction framework

Few-shot learning aims at enabling models to learn and generalize from only a few labeled data (Y. Wang et al., 2020). Unlike traditional supervised learning methods (e.g., GBDT), few-shot learning allows models quickly adapt to new tasks by leveraging pre-trained knowledge and minimal samples. Few-shot learning identifies a few representative samples as labeled samples from a sample dataset to form the so-called 'support set'. The rest of the samples in the dataset are classified as unlabeled samples that are included in the so-called 'query set'. The basic process of few-shot learning by prompt involves the following steps: firstly, the variables (the socioeconomic, built environment and other variables in our case) and the labels (the travel satisfaction variable in our case) of the samples in the support set are input into the LLM as prompts. Then, the unknown labels of the samples in the query set are predicted based on the known variables of these samples, which are input as prompts. The pipeline of the few-shot prediction framework in this study is outlined in Figure 2. Since the representativeness of samples in support set is reflected in the closeness to the characteristics and distribution of the target task (Y. Wang et al., 2020), the whole dataset is firstly divided into training set and test set through a process of random assignment, then the similarities between samples in these two sets are calculated. The top $k$ samples in the training set are selected to form the 'support set' used for the few-shot learning. After the LLM has sufficiently learned the latent patterns from these $k$ representative samples, we use the knowledge to predict the travel satisfaction of all remaining samples. The similarity between samples is calculated based on the Euclidean distance. The formula is as follows, in which $S_i$ is the $i$-th record, $S_i^k$ is the $k$-th feature of $i$-th record.

$$sim(S_i, S_j) = \frac{1}{\sqrt{\sum_{k=1}^{n}(S_i^k - S_j^k)^2 + 1}} \tag{1}$$

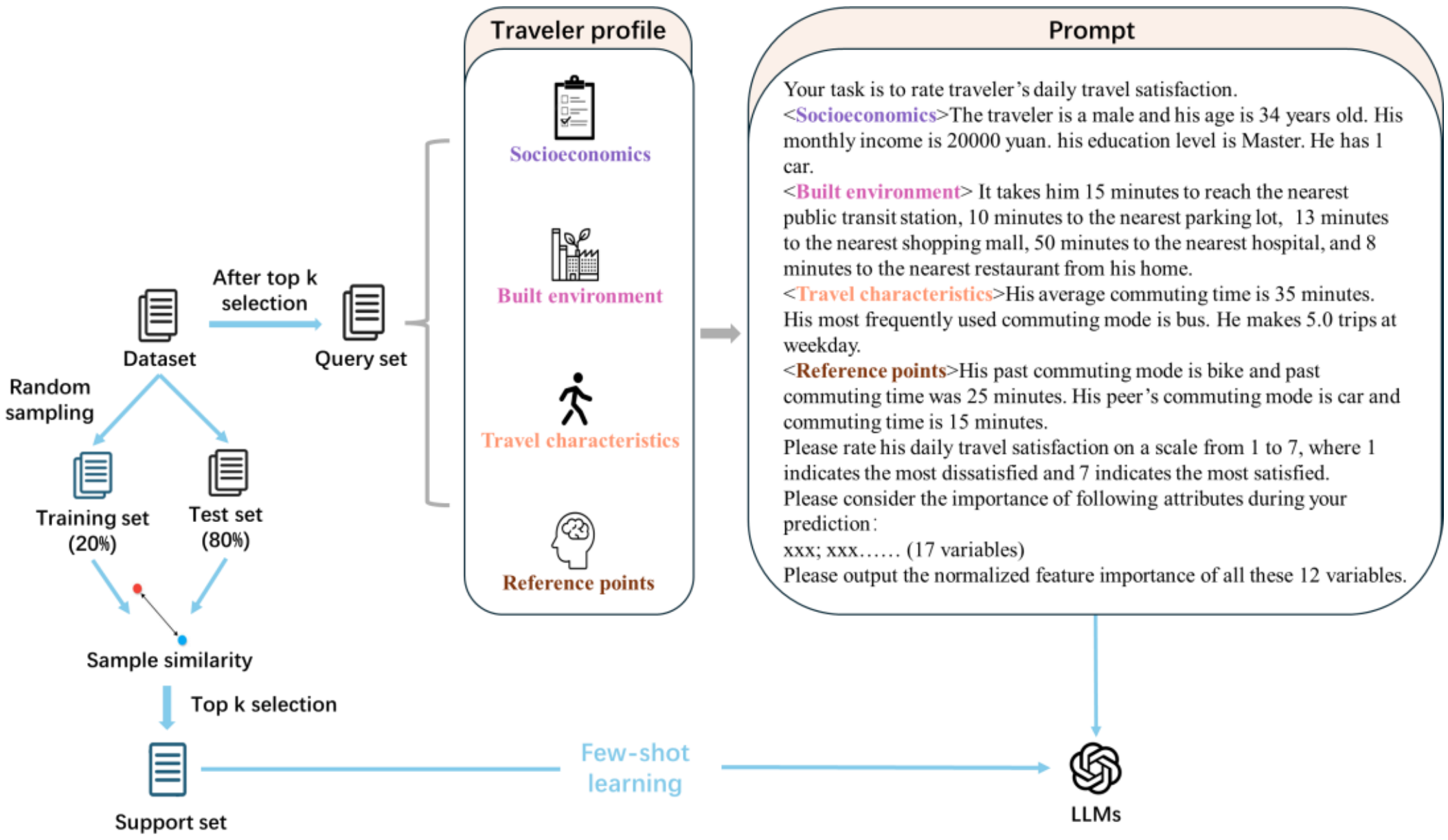

**Figure 2** Few-shot prediction framework

### *3.3 Experiment*

One of the most important hyperparameters of LLMs is temperature, typically ranging from 0 and 1. It controls the output probability distribution. A lower temperature generates stable output, while a higher temperature produces more creative output. To analyze the impact of temperature, we set the temperature at 0.2 and 0.7, respectively. LR and GBDT are representative statistical and machine learning models in travel satisfaction research. These two models are selected as baseline prediction models to assess the performance of LLMs. We evaluate the prediction accuracy of baseline models on the test set, the zero-shot LLM on the whole dataset, and the few-shot LLM on the query set, using four widely accepted metrics for model fit comparison: Mean Squared Error (MSE), Root Mean Squared Error (RMSE), Mean Absolute Error (MAE), and Mean Absolute Percentage Error (MAPE). All these metrics measure the difference between predicted and real values, thus the smaller they are, the better the model performance. Their mathematical expressions are shown below, in which $n$ is the number of samples, $y_i$ is the true value, and $\hat{y_i}$ is the predicted value. Each accuracy is computed with five repeated experiments.

$$MSE = \frac{1}{n}\sum_{i=1}^{n}(y_i - \hat{y_i})^2 \tag{2}$$

$$RMSE = \sqrt{\frac{1}{n}\sum_{i=1}^{n}(y_i - \hat{y_i})^2} \tag{3}$$

$$MAE=\frac{1}{n}\sum_{i=1}^{n}|y_i - \hat{y_i}| \tag{4}$$

$$MAPE = \frac{1}{n}\sum_{i=1}^{n}\frac{|y_i - \hat{y_i}|}{y_i} \tag{5}$$

## 4. Results

To evaluate the role of few-shot learning in aligning LLMs with human behavior, we shall compare the prediction accuracies of the few-shot LLM with those of zero-shot LLM and baseline models. Next, we will investigate the differences in the variable importance between GBDT and LLMs to identify the root causes of misalignment. Lastly, we aim to generalize the few-shot LLM framework to develop a small sample-based LLM modeling approach for travel behavior modeling.

### *4.1 The role of few-shot learning*

To demonstrate the effectiveness of few-shot learning, we systematically evaluate the prediction accuracies of the few-shot LLM using support sets of different sizes. Specifically, the size of the support set begins at zero, representing the zero-shot scenario, and increases in increments of three samples. The value of this increment does not have a uniform standard in the research related to few-shot learning. For instance, T. Liu et al. (2024) similarly used an increments of three samples. Pecher et al., 2024 conducted repeated experiments with increment of one sample. Shao et al., 2024 and Ye et al., 2022 even did not set a fixed value at all. The number of samples in the support set is typically a few or several dozen to fully demonstrates the low dependency of few-shot learning on sample size.

Table 3 and Table 4 lists the prediction accuracies of the LLMs using support sets of various sizes. The prediction accuracies of all LLMs improve with the aid of support set. This relationship between prediction accuracy and support set size aligns with the findings of Crulis et al., 2024 and Ye et al., 2022. They found that once the number of samples in the support set reaches a certain threshold, the marginal improvement in performance by few-shot learning gradually diminishes and may even become negative as the support set size continues to increase. Furthermore, the temperature hyperparameter appears to have little to no significant effect on the model's accuracy. These results answer the first and second research questions: the zero-shot LLM shows the existence of misalignment, causing a relatively low prediction accuracy; few-shot learning can reduce the misalignment and serves as an effective alignment method.

**Table 3** Prediction accuracies of the few-shot methods (temperature=0.2)

| Support set size | Metrics | DeepSeek-R1-671b | Qwen-3-235b | GPT-4o-200b | Llama3.3-70b |
|---|---|---|---|---|---|
| 0 (zero-shot) | MSE | 0.907 (0.045) | 3.657 (0.038) | 1.610 (0.035) | 2.406 (0.027) |
| | RMSE | 0.947 (0.024) | 1.912 (0.010) | 1.269 (0.014) | 1.550 (0.041) |
| | MAE | 0.768 (0.013) | 1.752 (0.013) | 1.060 (0.016) | 1.404 (0.033) |
| | MAPE | 0.190 (0.003) | 0.433 (0.004) | 0.261 (0.004) | 0.356 (0.017) |
| 3 | MSE | 0.701 (0.036) | 1.014 (0.078) | 1.066 (0.055) | 1.157 (0.056) |
| | RMSE | 0.837 (0.022) | 1.006 (0.039) | 0.987 (0.032) | 1.065 (0.049) |
| | MAE | 0.625 (0.024) | 0.803 (0.042) | 0.860 (0.069) | 0.855 (0.072) |
| | MAPE | 0.146 (0.010) | 0.170 (0.001) | 0.199 (0.004) | 0.207 (0.052) |
| 6 | MSE | 0.752 (0.022) | 0.817 (0.050) | 0.909 (0.008) | 1.149 (0.051) |
| | RMSE | 0.866 (0.047) | 0.903 (0.028) | 0.953 (0.004) | 1.058 (0.096) |
| | MAE | 0.634 (0.030) | 0.688 (0.050) | 0.733 (0.022) | 0.821 (0.028) |
| | MAPE | 0.147 (0.010) | 0.160 (0.019) | 0.175 (0.008) | 0.199 (0.009) |
| 9 | MSE | 0.676 (0.057) | 0.698 (0.063) | 0.896 (0.059) | 1.227 (0.053) |
| | RMSE | 0.820 | 0.835 | 0.940 | 1.099 |
| | | (0.063) | (0.037) | (0.029) | (0.059) |
| | MAE | 0.604 (0.071) | 0.606 (0.033) | 0.693 (0.055) | 0.880 (0.078) |
| | MAPE | 0.142 (0.021) | 0.139 (0.007) | 0.161 (0.045) | 0.216 (0.47) |
| 12 | MSE | 0.701 (0.029) | 0.716 (0.039) | 0.895 (0.080) | 1.187 (0.025) |
| | RMSE | 0.837 (0.017) | 0.846 (0.023) | 0.932 (0.029) | 1.086 (0.034) |
| | MAE | 0.595 (0.022) | 0.626 (0.022) | 0.702 (0.024) | 0. 825 (0.030) |
| | MAPE | 0.133 (0.004) | 0.143 (0.005) | 0.167 (0.003) | 0. 199 (0.012) |

*Notes*. Underlined numbers indicate the highest accuracy. Italic numbers in the bracket are the standard deviations of repeated experiments. The highest accuracy is highlighted in blue.

**Table 4** Prediction accuracies of few-shot learning (temperature=0.7)

| Support set size | Metrics | DeepSeek-R1-671b | Qwen3-235B | GPT-4o-200b | Llama3.3-70b |
|---|---|---|---|---|---|
| 0 (zero-shot) | MSE | 0.932 (0.043) | 3.592 (0.069) | 1.656 (0.031) | 2.415 (0.033) |
| | RMSE | 0.965 (0.022) | 1.895 (0.018) | 1.287 (0.012) | 1.553 (0.043) |
| | MAE | 0.771 (0.022) | 1.729 (0.021) | 1.086 (0.013) | 1.407 (0.024) |
| | MAPE | 0.191 (0.005) | 0.428 (0.007) | 0.268 (0.003) | 0.357 (0.008) |
| 3 | MSE | 0.759 (0.023) | 1.365 (0.038) | 1.049 (0.052) | 1.126 (0.083) |
| | RMSE | 0.871 (0.038) | 1.163 (0.027) | 1.006 (0.047) | 1.061 (0.040) |
| | MAE | 0.634 (0.083) | 0.989 (0.034) | 0.780 (0.076) | 0.845 (0.050) |
| | MAPE | 0.147 (0.029) | 0.245 (0.036) | 0.187 (0.049) | 0.204 (0.016) |
| 6 | MSE | 0.769 (0.016) | 0.842. (0.086) | 1.259 (0.082) | 1.320 (0.090) |
| | RMSE | 0.875 (0.015) | 0.913 (0.051) | 1.087 (0.032) | 1.144 (0.024) |
| | MAE | 0.672 (0.011) | 0.662 (0.050) | 0.850 (0.053) | 0.919 (0.067) |
| | MAPE | 0.159 (0.020) | 0.151 (0.028) | 0.200 (0.034) | 0.220 (0.049) |
| 9 | MSE | 0.666 (0.023) | 0.724 (0.017) | 0.972 (0.062) | 1.209 (0.055) |
| | RMSE | 0.815 (0.013) | 0.849 (0.068) | 0.986 (0.031) | 1.088 (0.041) |
| | MAE | 0.604 (0.021) | 0.611 (0.044) | 0.746 (0.046) | 0.842 (0.032) |
| | MAPE | 0.139 (0.004) | 0.136 (0.008) | 0.176 (0.017) | 0.220 (0.009) |
| 12 | MSE | 0.670 (0.031) | 0.741 (0.071) | 0.830 (0.070) | 1.285 (0.073) |

| | | | | |
|---|---|---|---|---|
| RMSE | 0.819 (0.019) | 0.860 (0.042) | 0.908 (0.036) | 1.124 (0.038) |
| MAE | 0.614 (0.036) | 0.599 (0.031) | 0.654 (0.038) | 0.898 (0.017) |
| MAPE | 0.147 (0.011) | 0.136 (0.007) | 0.151 (0.010) | 0.217 (0.014) |

As shown in Figure 3 and Figure 4, the baseline models with best performance are compared with few-shot LLMs that achieve the highest prediction accuracy. The prediction errors of the linear regression model reaches its lowest with a MAPE of 0.171 when 90% of the samples are used to develop the model. The highest accuracy of the GBDT model is observed when 80% of the samples are used, with a MAPE of 0.183. It is worth noting that, unlike LR, GBDT models have some degree of randomness in their parameters, which means that the highest accuracy may not necessarily be achieved with the largest percentage of samples. All LLMs, except for Llama, have outperformed baseline models, demonstrating the strong potential of LLMs for travel satisfaction modeling.

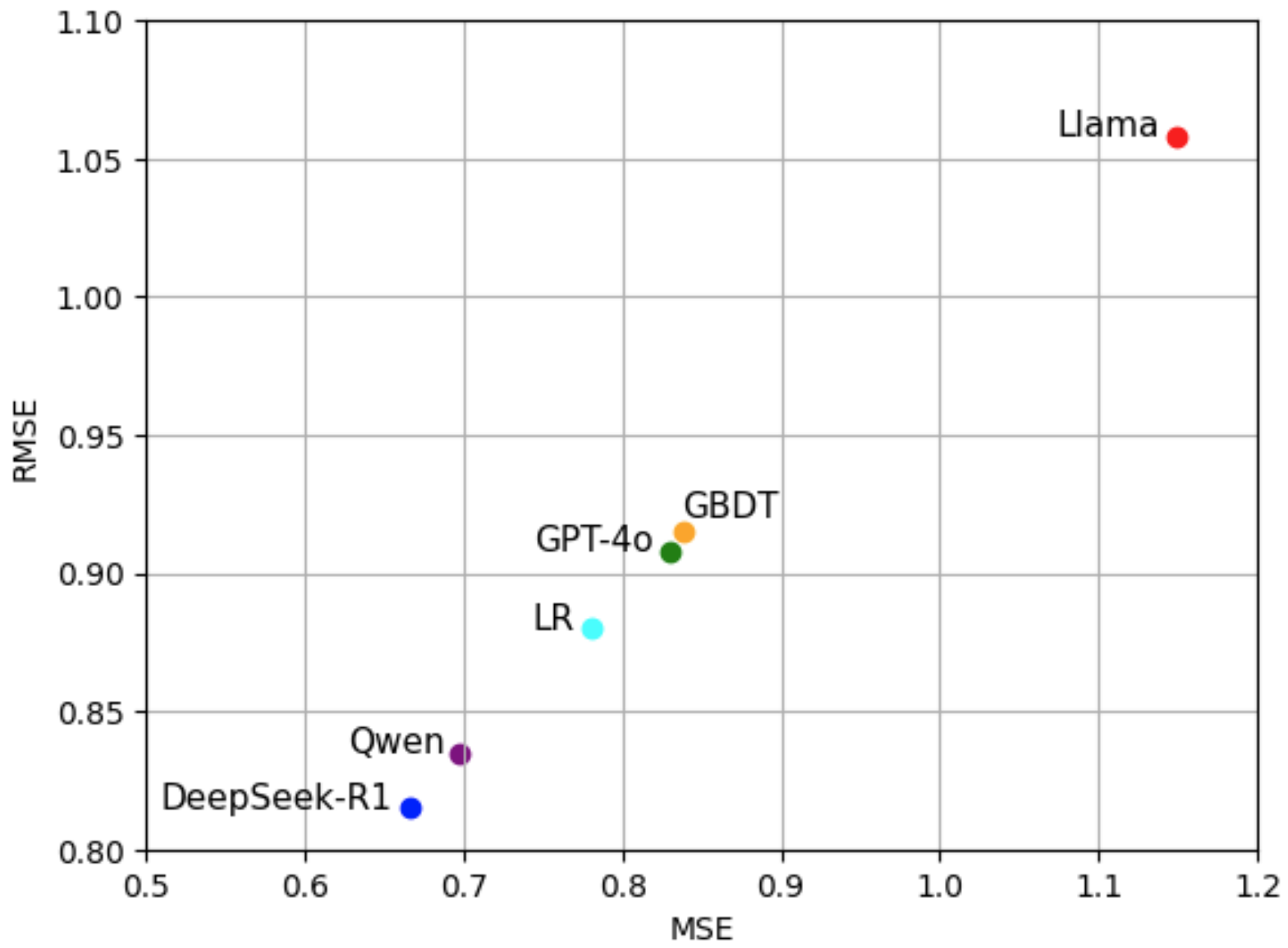

**Figure 3** MSE and RMSE of baseline models and LLMs

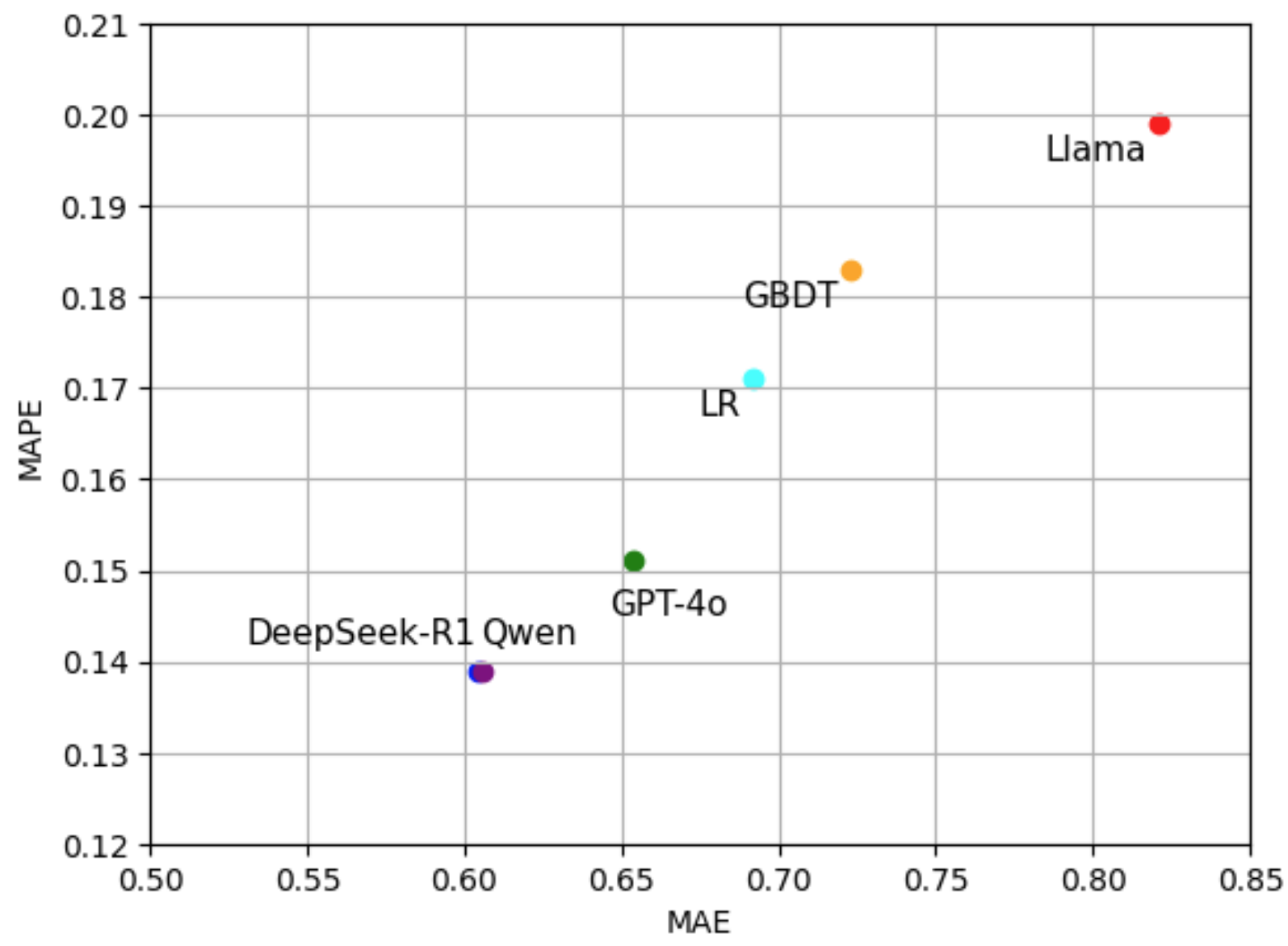

**Figure 4** MAE and MAPE of baseline models and LLMs

### *4.2 Variable importance in LLMs and GBDT*

Both the zero-shot LLM and the few-shot LLM were requested to provide normalized importances it assigned to variables. Figure 5 and Figure A1 to A3 (see Appendix) compare the average importance of variables provided by GBDT, the zero-shot LLM, and the few-shot LLM achieving the highest accuracy. The variable importance distributions across the four LLMs show similar patterns. For further analysis, we focus on Deepseek-R1, which demonstrates the highest accuracy along with high stability in its results.

On the one hand, the importance of most variables in GBDT exhibits obvious differences compared to those in both LLMs. Commuting time and transport mode are found to have a more substantial impact in LLMs than in GBDT. In contrast, the number of trips on weekday is almost negligible in GBDT, and similarly assigned a low importance in LLMs. The built environment variables generally play more important roles in GBDT than in LLMs. Regarding socioeconomic variables, aside from car access, the others hold little importance across all three models. Car access is highly weighted in the zero-shot LLM, but its importance decreases in the few-shot LLM. Concerning the reference point dimension, past commuting time is the most important variable in GBDT but is assigned a low weight in both zero-shot and few-shot LLMs. Similar contrasts are observed for peer’s commuting mode. Past commuting mode has almost the same importance in all these models. Peer’s commuting time holds limited importance in GBDT but emerges as the fifth most important variable in the few-shot LLM. These differences among models may explain the models' varying prediction accuracies. Although GBDT and LLMs show great differences in their assessment of variable importance, they exhibit close levels of prediction accuracy, which may suggest that they follow different mechanisms for modeling travel satisfaction. On the other hand, the differences in variable importance between the zero-shot LLM and the few-shot LLM may show the impact of aligning process. For example, the importance of car access, number of trips, and past commuting mode in the few-shot LLM are lower than those in the zero-shot LLM, which may explain why the few-shot LLM has a better predictive performance.

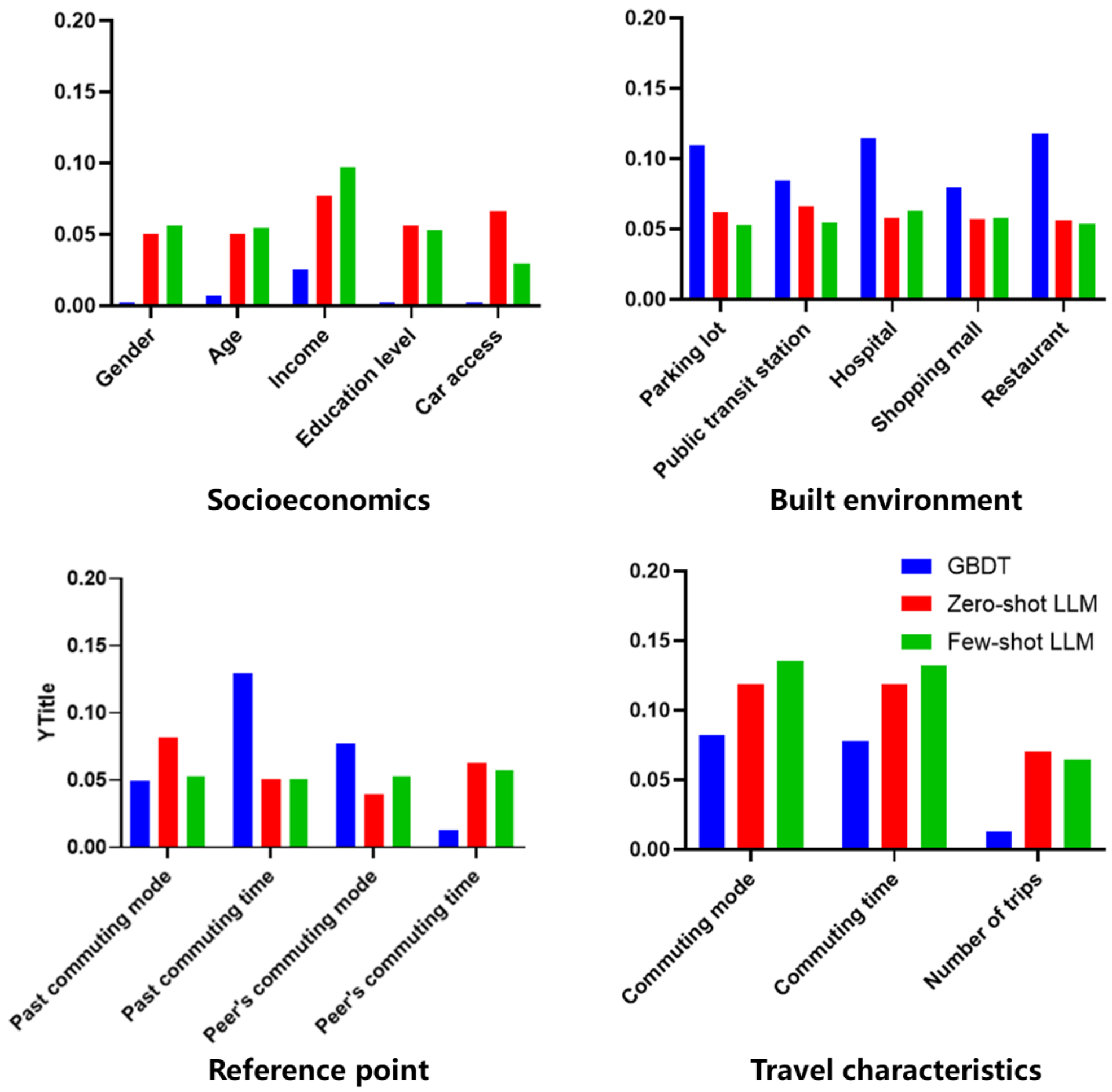

**Figure 5** Variable importance in GBDT and DeepSeek-R1

Although we have identified differences in variable importance between the zero-shot

LLM and the few-shot LLM, it remains unclear how these differences come from. In other words, how the aligning process shapes the LLMs' behavior are not yet understood. DeepSeek-R1's ability to generate its thinking process before generating the output offers a rational and feasible approach to addressing this problem. We find that the model automatically categorize variables into three groups when determining their importances—low-weight, middle weight, and high weight—and give detailed explanations (see Table A1 in Appendix). Through the deep-thinking contents of the zero-shot LLM and the few-shot LLM summarized in Table 5, more valuable insights into the aligning process can be obtained.

The zero-shot LLM's reasoning is primarily developed during the pre-training stage using both quantitative and qualitative data related to travel behavior. On the other hand, the reasoning of the few-shot LLM incorporates the information provided by the samples in the support set. For example, the zero-shot LLM assigns a high weight to car access, reasoning that the number of cars determines the flexibility and comfort of travel. However, the samples in the support set all have the same value for this variable, providing no meaningful statistical information for predicting travel satisfaction. As a result, the weight of this variable is reduced in the few-shot scenario. Similarly, the zero-shot LLM assumes that more trips lead to greater fatigue, which aligns with common sense. However, the few-shot LLM revises this judgment and reduces the weight of this variable.

For variables that have similar weights in both zero-shot and few-shot LLMs, the interpretations may vary. For example, although commuting mode and time are assigned with high weight, the interpretations by the zero-shot LLM are likely drawn from the relevant literature, whilst those by the few-shot LLM are clearly influenced by the training data provided. The zero-shot LLM believes that, aside from car access, variables on socioeconomics can influence travel satisfaction through indirectly influencing preferences. It is in line with many previous studies pointing out that socioeconomic variables are less predictive of travel satisfaction (Bergstad et al., 2011; Humagain and Singleton, 2020; Olsson et al., 2013). After the few-shot learning, the LLM finds that the impact of these variables is not significant in support set and thus assigns them a low weight as well. Although the weights of the built environment variables are consistent across the models, LLMs categorize them into two groups:

accessibility to transit stations and parking lots, and accessibility to other living facilities. It concludes that the former directly affects the travel satisfaction of both public transit commuters and car commuters, hence assigning a middle weight to it. The latter mainly impacts the convenience of daily life and has a lower influence on travel satisfaction. The finding may resonate with some studies on the built environment, which has consistently demonstrated that distance to public transit and parking spaces significantly influence travel mode choice (Christiansen et al., 2017; De Gruyter et al., 2020). This result highlights the necessity of developing separate models for different travel modes when examining the impact of the built environment on travel satisfaction.

In both LLMs, past travel experience and peer's travel are assigned middle weights. The zero-shot LLM deems that past travel experience serves as a reference for current choices. However, after learning from the support set, although the weights remain unchanged, the few-shot LLM holds the opinion that individual differences play a key role. Similarly, peer's travel is assigned a middle weight in both LLMs due to its social impact on individuals' choices. These results are strongly supported by the Prospect Theory, which proposes that past performance (i. e., past travel experience) serves as a non-status quo reference point that affects current expectations, and social comparison (i. e., peer's travel) is a crucial reference point for individual self-evaluation (Kahneman and Tversky, 2013). The differences in the reasoning contents of LLMs fundamentally clarify the pivotal role of few-shot learning in the aligning process.

While the importance of some variables remains relatively stable, the comprehension of these variables by LLMs has evolved considerably. The progression from the zero-shot LLM to the few-shot LLM signifies a transition from broad, general reasoning to domain-specific understanding. We conclude that the misalignment of LLMs stems from the discordance between the general knowledge of pre-trained LLMs and the unique characteristics of the dataset. The superior predictive performance of the few-shot LLM can be attributed to the good integration of these two aspects.

**Table 5** Deep-thinking contents during reasoning in zero-shot and few-shot LLMs

| Variables | Zero-shot | Few-shot |
|---|---|---|
| Time to nearest transit station and parking lot | Middle weight, as proximity to transit stations and parking lots enhances satisfaction especially for car and public transit commuters. | |
| Time to nearest hospital, restaurant, and mall | Low weight, as they are the accessibility of daily convenience facilities. | |
| Income | Low weight, as they may indirectly affect personal preferences, their weight may be lower. | Low weight, as there is no significant difference in satisfaction among high-income individuals in the samples. |
| Age | | Low weight, as the age distribution is wide but does not significantly affect satisfaction. |
| Gender<br>Education | | Low weight, as there is no obvious correlation in the samples. |
| Car access | High weight, as it determines whether a car is available, which affects the commuting flexibility and comfort. | Low weight, as almost all samples have no car, which may have no impact. |
| Commuting mode and time | High weight, as commuting time and mode are often considered key factors affecting commuting satisfaction. | High weight, shorter commuting times and active modes increase satisfaction. |
| Number of trips on weekday | Middle weight, as frequent trips may lead to more fatigue. | Low weight, as sample data is insufficient though fatigue reduces satisfaction. |
| Past commuting mode and time | Middle weight, as previous commuting methods and time may affect current choices. | Middle weight, as switching from long-term public transit or car to cycling may affect satisfaction, but the results may vary depending on individual differences. |
| Peer's commuting mode and time | Middle weight, as they may have a social impact on individuals' choices. | |

### *4.3 Generalizability of the LLMs modeling approach*

The finding that few-shot LLMs achieve prediction accuracy comparable to conventional statistical/econometric and GBDT models (as demonstrated in Section 4.1) suggests significant potential for applying LLMs in future travel behavior modeling. Depending on the number of variables involved, the commonly used statistical models such as linear regression and structural equations models often need sample sizes of at least hundreds to ensure models' statistical validity, according to the G-power test (Faul et al., 2007; Soper, 2016). Hoffmann et al. (2017), in a systematic review about travel mode choice modeling, recommended that a sample size greater than 200 or a power analysis of sample size were conducted as two inclusion criteria for studies based on statistical models. Apart from sample size, statistical models require a rigid and well-designed sampling strategy (e.g., stratified sampling method with PPS) to collect data. In other words, collecting data is usually of a high cost of both money and time, which is one of the key factors impeding the development of travel behavior modeling. On the other hand, few-shot LLMs require only a handful of samples, suggesting significantly lower data collection costs. However, in the few-shot framework illustrated in Figure 2, the representative samples in the support set are chosen from a large sample. This raises a crucial question: Is it necessary to have a large sample to select representative samples for the support set? If not, few-shot LLMs could be developed from a very small sample, which will drastically reduce the costs of developing travel behavior models. In the remainder of this subsection, we will address this important issue.

Recall that as illustrated in Figure 2 in Section 3.2, a key step in the few-shot LLM modeling is to identify representative samples to be included in the support set and such samples should be representative of the large sample set. Instead of selecting representative samples based on their similarity with the large sample set, we randomly select these samples to form the support set, and modify the few-shot LLM modeling approach to that shown in Figure 6. Such random selection mimics collecting data through random sampling in the field. To demonstrate that these samples are representative of the overall dataset and can thus serve as the support set in few-shot learning, we conduct Kolmogorov-Smirnov tests (K-S test) for each variable. These tests are used to determine whether there are significant differences between the support set and the full dataset. If no significant difference is detected, it suggests that the support set can represent the overall dataset. The results are listed in Table 6. As one tell from the table, in most cases, the distributions of variables from the randomly sampled subset have no significant differences from the overall distribution. More crucially, despite the increase of prediction instability caused by randomization. the few-shot LLM still exhibits a remarkably superior performance. This finding aligns with the study made by Hagendorff et al. (2023), who found that GPT-3's ability to answer correctly increased with randomly-selected additional training examples that the task was prefixed with.

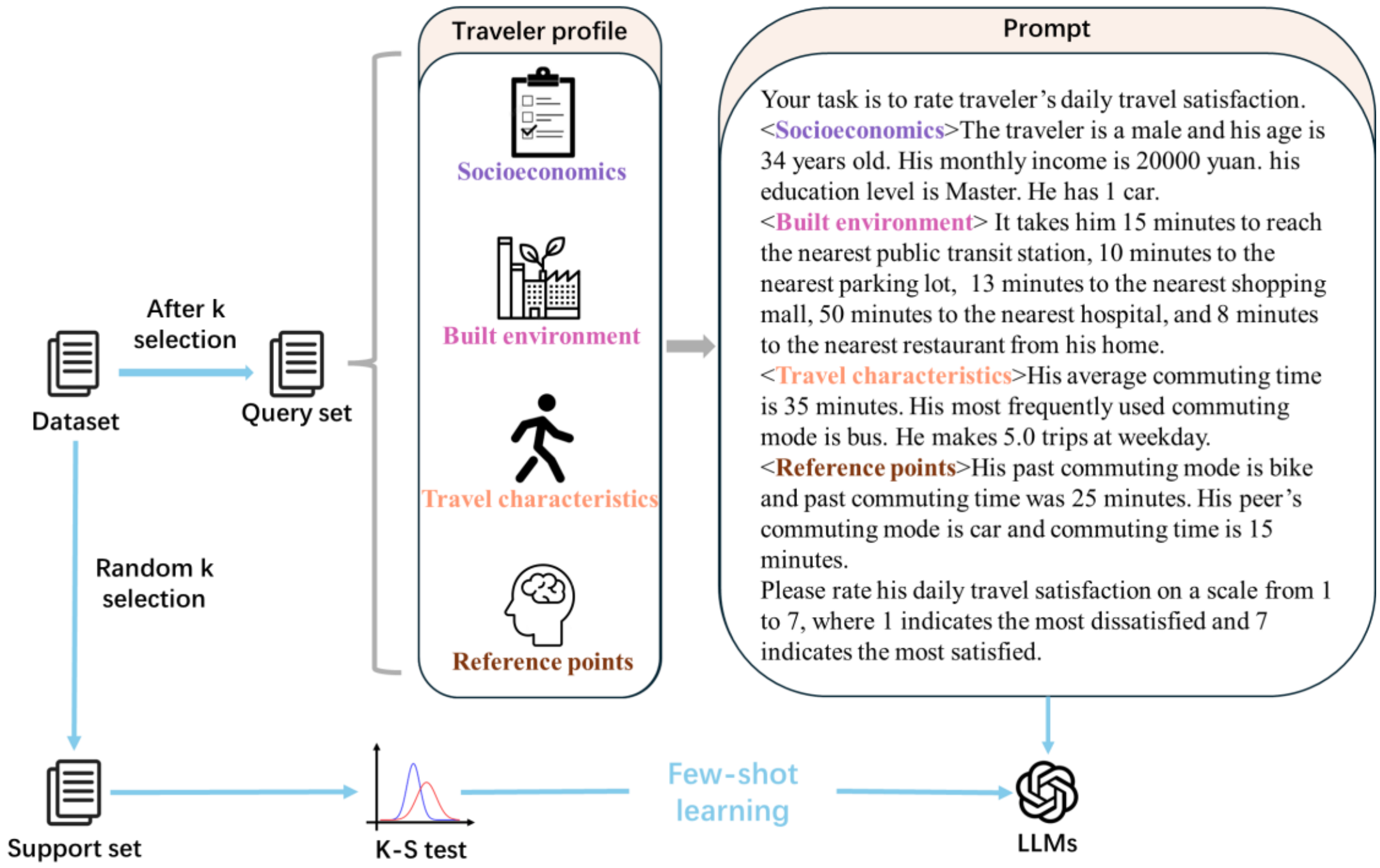

**Figure 6** A simplified few-shot prediction framework

**Table 6** The K-S test results and prediction accuracy of DeepSeek-R1 using random sampling (temperature=0.7)

| Support set size | K-S test | MSE | RMSE | MAE | MAPE |
|---|---|---|---|---|---|
| 3 | ns | 0.736 (0.046) | 0.835 (0.032) | 0.660 (0.021) | 0.157 (0.006) |
| 6 | Public transit station* (1) | 0.730 (0.053) | 0.851 (0.021) | 0.655 (0.032) | 0.160 (0.005) |
| 9 | Parking lot* (1) | 0.716 (0.087) | 0.843 (0.038) | 0.632 (0.022) | 0.154 (0.002) |
| 12 | Shopping mall* (1) | 0.665 (0.035) | 0.810 (0.020) | 0.597 (0.023) | 0.142 (0.006) |

*Notes.* The numbers in bracket in the K-S test column represent the number of times that significant differences appear in five repeated samplings. Significance at 0.05 and 0.01 level are marked with * and ** respectively. "ns" means there is no significant difference in all repeated samplings.

The above results reveal that even a randomly selected support set can be used to align LLMs. In other words, by employing standard sampling design strategies to ensure the representativeness of samples (e.g., socioeconomic and geographical representativeness), we can construct a support set through collecting data from a relatively small sample, instead of identifying samples from a large sample. This conclusion is strongly supported by Zhou et al., 2023, as they have proved that the alignment of LLMs can be elicited through a handful of carefully curated examples. Building on this, we propose an LLM-based modeling approach as shown in Figure 7, which can be applied to modeling travel behavior in general using small samples. Apart from travel satisfaction, other dimensions of travel behavior such as choices of transport mode, activity destination, and trip purpose can be modeled using this approach (e. g., Liang et al., 2024; Shao et al., 2024).

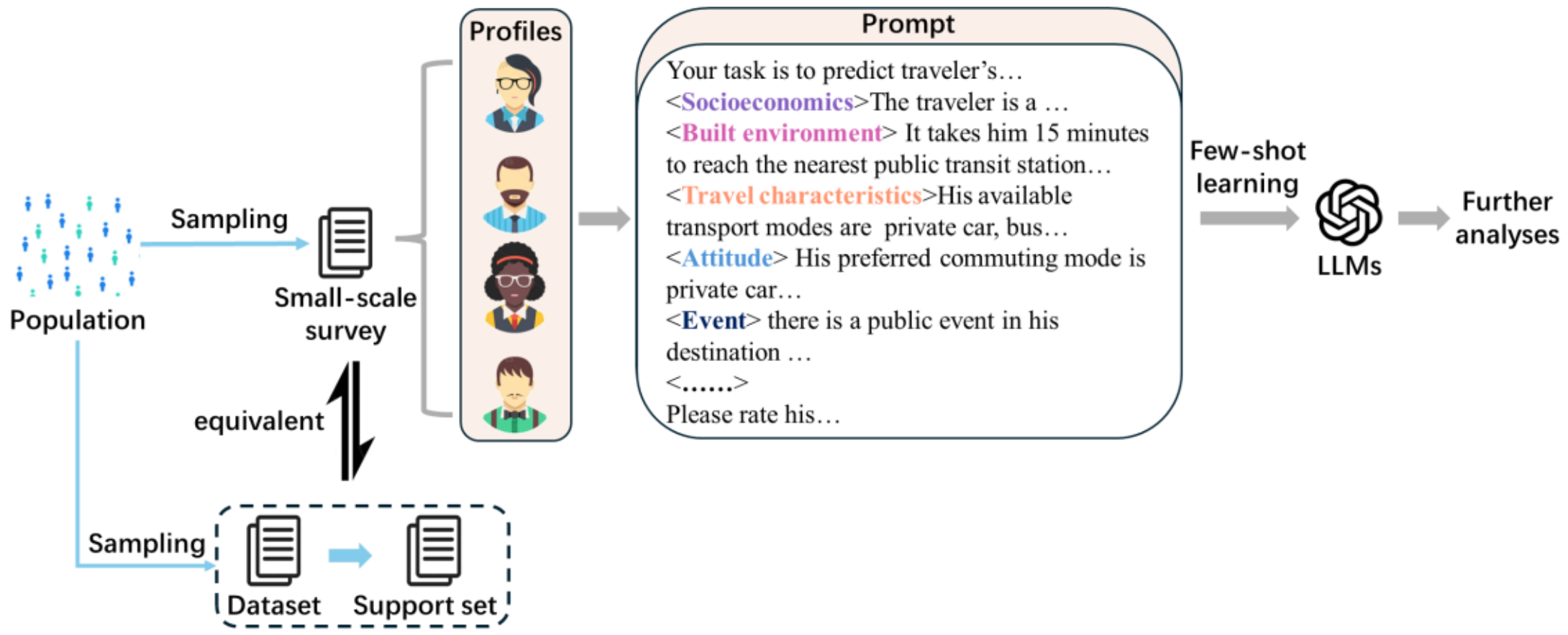

**Figure 7** A few-shot LLM-based travel behavior modeling approach

## 5. Discussions and conclusions

This study investigates the application of LLMs in travel satisfaction research and answers three research questions: (1) whether there is a misalignment between LLMs and human travelers on travel satisfaction? (2) how to improve their alignment? and (3) what are the sources of the misalignment? The research is motivated by the broad application prospects and potential risk of LLMs in travel behavior modeling. The determinants of travel satisfaction are obtained and incorporated into prompts based on data from a household travel survey conducted in Shanghai. We employed both zero-shot and few-shot LLMs to model and predict travel satisfaction. Their performances are compared with that of linear regression and GBDT models, which are treated as baseline models.

Through detailed analyses of the prediction results and variable importances, we have following research findings. Concerning the first question, this study finds an obvious misalignment between LLMs and human travelers, resulting in low prediction accuracy of zero-shot LLM. For example, car access is an important variable to predict travel satisfaction in zero-shot LLM, whereas it is assigned a low weight in both GBDT and few-shot LLM. For the second question, this study proves that few-shot learning can efficiently improve the alignment using only minimal labeled samples, making the prediction accuracy outperforms the baseline models. Some studies have pointed out that an essential requirement for applying LLMs is lifelong learning or continuous learning, which emphasizes the models' ability to update their knowledge based on new data or information (Li et al., 2023). This viewpoint appears to resonate with our findings: by introducing a small number of labeled samples into the LLM, its understanding of the factors affecting travel satisfaction is improved and hence the accuracy of travel satisfaction prediction is also upgraded. Regarding the third question, we revealed significant differences in variable importance among the GBDT model, zero-shot, and few-shot LLMs. The reasoning contents show discrepancies between the zero-shot LLM, which is based on the general knowledge about travel satisfaction from pre-trained LLMs, and the few-shot LLM that is equipped with domain-specific knowledge from samples in the support set. Such discrepancies are likely the major sources of the LLMs' misalignment.

Based on these research findings, we further explore the potential of the few-shot LLM modeling framework for travel behavior modeling in general. We argue that LLMs hold the potential to revolutionize the paradigm of travel behavior modeling that heavily relies on questionnaire survey data. Conducting questionnaire surveys demands substantial manpower and financial commitments, which makes repeated and timely data collections almost impossible. This study reveals that LLMs can generate accurate estimations and give interpretable insights into the modeling process with a sample of very small size. This capability suggests that LLMs hold great potentials for substantially reducing the costs for collecting data to develop travel behavior models.

Despite the advantages of LLMs, we find that they are prone to hallucination, high cost, and inefficiency as well, which may hinder their broader adoption. For example, when LLMs output the travel satisfaction scores of a batch of travelers, repetitive patterns often emerge in the results. This is referred to as the "parroting problem" of LLMs, which is caused by greedy strategy and self-reinforcement effect (Xu et al., 2022). To address this issue, this study adopts two strategies. Additionally, since the designed prompt contains many variables and experiments are repeated five times, modeling development cost is relatively high. Apart from monetary costs, there are also time costs. As the expenditure of deploying complete LLMs locally is unaffordable, we use remote API to access the commercial LLMs platform, which are plagued by response inefficiency. The resolutions to high cost and inefficiency depend on the development of open-source LLMs and the improvement of AI infrastructure, whilst the hallucination can be partially addressed through prompt engineering, but fundamentally relies on the advancement of LLM technology.

This study can be further improved in the following directions. First, this study makes use of sample survey data collected in Shanghai, China to evaluate the prediction capability of LLMs. Data collected in other places can be used to verify the research findings of this study. Second, since LLMs' performance is highly related to the number of parameters, the external validity of few-shot learning can be further validated using distilled models or other series of LLMs. Third, the discussions on misalignment between LLMs and human travelers can be extended to other dimensions of travel behavior such as travel mode choice, activity type prediction, and activity destination choice, etc.

## Appendix

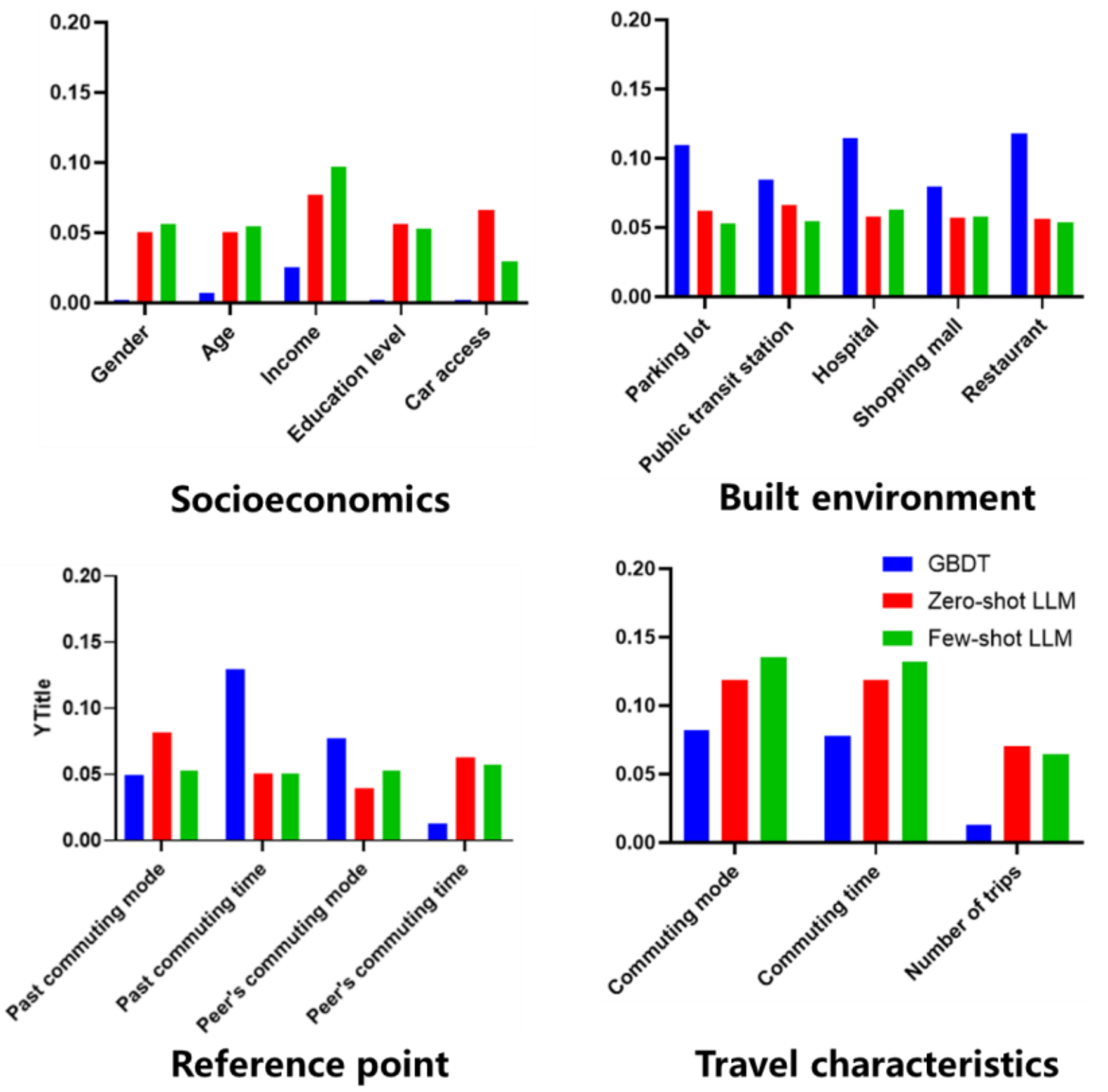

**Figure A 1** Variable importance in GBDT and GPT-4o

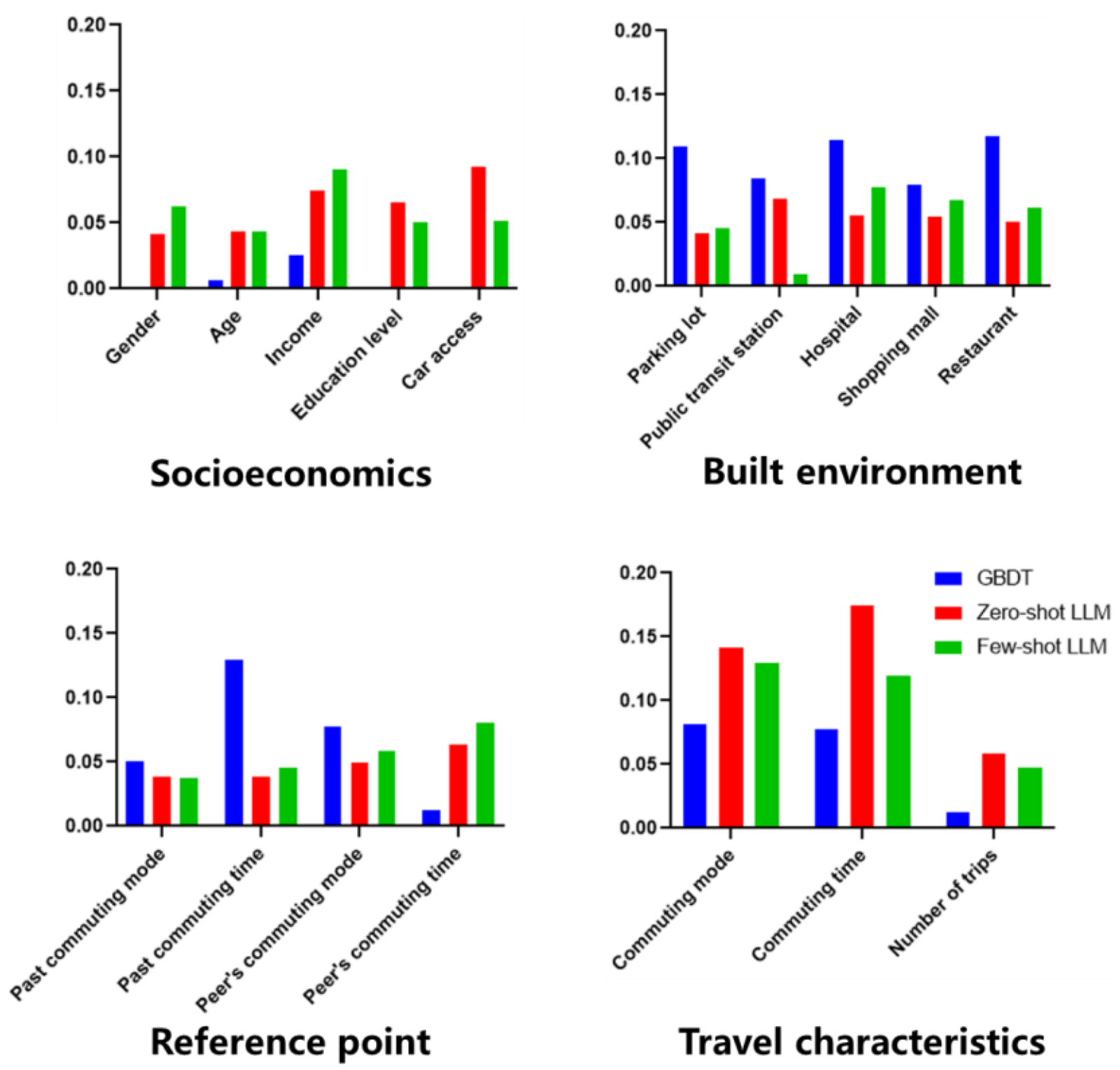

**Figure A 2** Variable importance in GBDT and GPT-4o

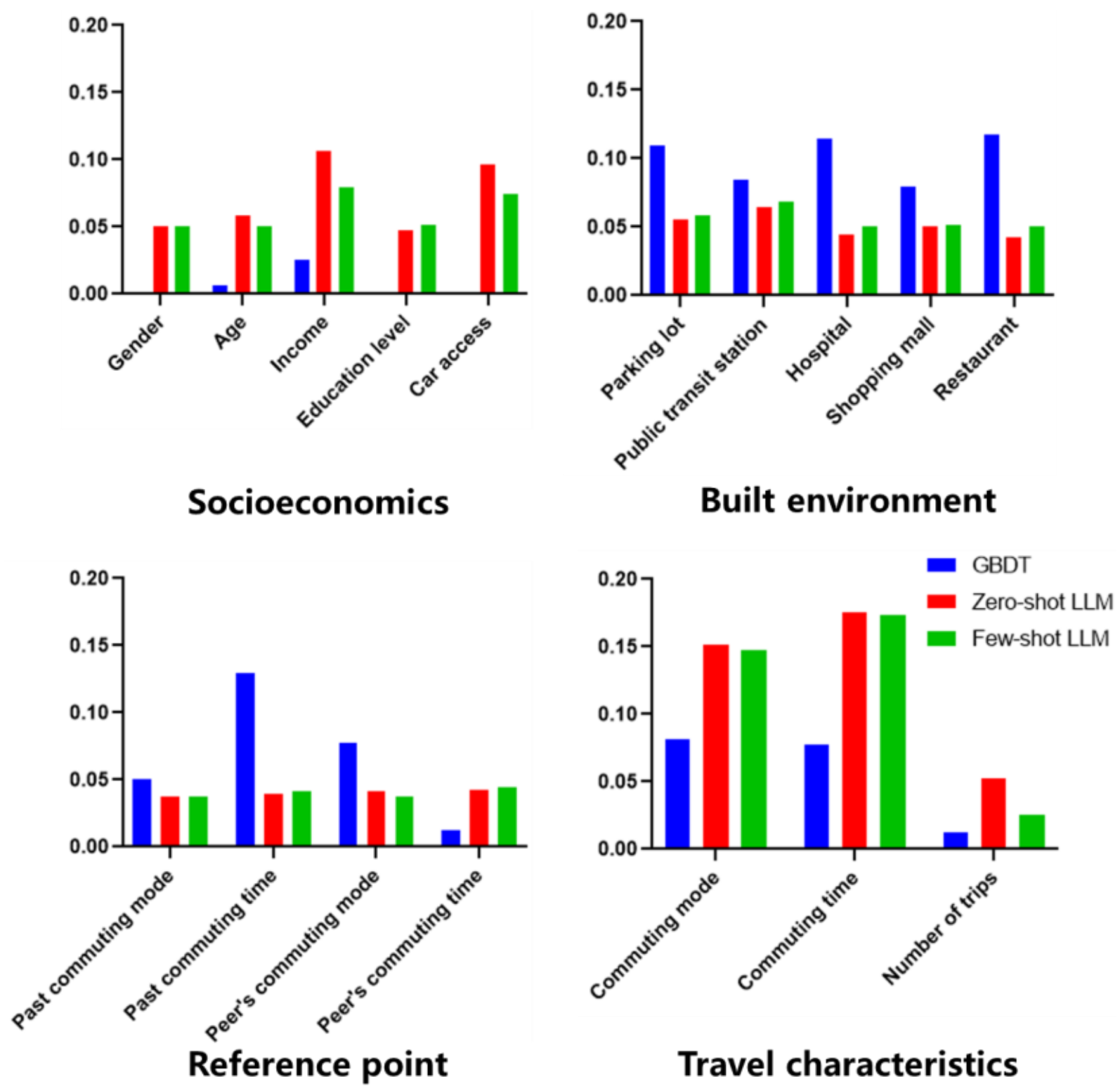

**Figure A 3** Variable importance in GBDT and Qwen

**Table A 1** An example of the deep-thinking content produced by DeepSeek-R1 during the experiment

| DeeSeek-R1-671b |
| --- |
| First, I need to clarify the users' needs. They may want to know which factors have the greatest impact on travel satisfaction, or they may want to build a model to predict satisfaction and need to know the weights of each feature. The users could be researchers, data analysts, or practitioners in transportation planning, hoping to identify key influencing factors through these data and then develop improvement measures.<br>Next, I need to consider how to determine the weights of each element. Typically, this can be done through statistical analysis methods, such as regression analysis (linear regression or logistic regression), to calculate the coefficients of each variable. The size and sign of the coefficients can indicate the direction and weight of the impact. However, the users have not provided the original dataset but only a textual description, so actual statistical analysis cannot be performed. At this point, I need to conduct an observational analysis based on the provided samples, identify possible patterns and correlations, and thereby infer the relative importance of each element.<br>Short commuting time, using walking or bicycles, and proximity to surrounding facilities are associated with higher satisfaction, while using company shuttles, long commuting time, and distant facilities are associated with lower satisfaction. Meanwhile, income and education level do not seem to have a clear direct correlation, but there may be indirect effects. For example, high-income individuals may prefer to use private cars, but they may have longer commuting times, which in turn affects satisfaction.<br>Now, I need to transform these observations into estimates of weights. Weights usually indicate the degree of influence of a variable on the target variable (satisfaction). Due to the |

lack of data, I need to make reasonable inferences based on the trends in the sample and clearly state that this is an assumption based on observation, not rigorous statistical analysis. Next, I need to list all possible elements and assign a weight to each one. For example:
1. Commuting time: Likely to have a high weight, as multiple samples show that short commuting time is associated with high satisfaction.
2. Commuting mode: Different modes have different impacts. For example, walking/bicycling may have a high weight, while company shuttles may have a low weight. Distance to surrounding facilities: Such as the distance to public transportation stations, hospitals, and shopping centers, likely to have a medium weight.
3. Income: Likely to have a low weight, as high-income individuals in the sample show a wide range of satisfaction levels.
4. Age: Likely to have a low weight, as the age distribution is broad but does not significantly affect satisfaction.
5. Peer's commuting mode: There may be a social comparison effect, so it could have a medium weight.
6. Previous commuting mode: Changes may affect satisfaction, but the weight is likely to be low.
7. Education level: No clear correlation in the sample, so it has a low weight.
8. Number of cars: All samples have one vehicle, so it may have no impact and a weight of zero.
(……)
It is important to note that some factors may have interaction effects. For example, commuting mode and commuting time may jointly affect satisfaction, but they need to be considered separately when analyzed individually. Moreover, the convenience of surrounding facilities may indirectly affect the choice of commuting mode, which in turn affects satisfaction. It is also necessary to point out that these weights are based on observations from a limited sample and need further verification in practical applications.

## CRediT authorship contribution statement

**Pengfei Xu**: Conceptualization, Methodology, Formal analysis, Investigation, Writing – original draft. **Donggen Wang:** Resources, Validation, Supervision, Writing – review & editing.

## Declaration of Competing Interest

The authors declare that they have no known competing financial interests or personal relationships that could have appeared to influence the work reported in this paper.

## Acknowledgements

This research is supported by the following research grant: General Research Fund (GRF) grants from the Hong Kong Research Grant Council (HKBU12613324).